\documentclass[doublecol]{epl2} 

\usepackage{hyperref}
\usepackage{amssymb}
\usepackage{latexsym}
\usepackage{amsmath}

\newcommand{\myi}{$\acute\imath$}
\newcommand{\ds}{\displaystyle}
\newcommand{\G}{\mit{\Gamma}}

\renewcommand{\d}{\delta}
\renewcommand{\a}{\alpha}
\renewcommand{\b}{\beta}
\newcommand{\e}{\epsilon}
\newcommand{\f}{\phi}

\renewcommand{\l}{\lambda}
\newcommand{\g}{\gamma}
\renewcommand{\e}{\epsilon}
\newcommand{\m}{\mu}
\newcommand{\n}{\nu}
\renewcommand{\r}{\rho}
\newcommand{\s}{\sigma}

\renewcommand{\o}{\omega}

\newcommand{\pa}{\partial}

\renewcommand{\f}{\phi}

\newcommand{\uto}{\xrightarrow}

\newcommand{\realni}{\ensuremath{\mathbb{R}}}
\newcommand{\kompleksni}{\ensuremath{\mathbb{C}}}
\newcommand{\R}{{\realni}}
\newcommand{\C}{{\kompleksni}}

\newcommand{\rmd}{{\rm d}}
\newcommand{\cM}{{\cal M}}

\title{Standard Model and 4-groups}
\shorttitle{Standard Model and 4-groups} 

\author{A. Mikovi\'c\inst{1,2,*} \and M. Vojinovi\'c\inst{3,\dag}}
\shortauthor{A. Mikovi\'c \etal}

\institute{                    
  \inst{1} Departamento de Matem\'atica e COPELABS, Universidade Lus\'ofona de Humanidades e Tecnologias, Av. do Campo Grande, 376, 1749-024 Lisboa, Portugal\\
  \inst{2} Grupo de F\myi sica Matem\'atica da Universidade de Lisboa, Faculdade de Ci\^encias -- Campo Grande, Edif\myi cio C6, PT-1749-016 Lisboa, Portugal\\
  \inst{3} Institute of Physics, University of Belgrade -- Pregrevica 118, 11080 Belgrade, Serbia\\
  \inst{*} {\rm e-mail: amikovic@ulusofona.pt}\\
  \inst{\dag} {\rm e-mail: vmarko@ipb.ac.rs}
}

\abstract{
We show that a categorical generalization of the the Poincar\'e symmetry which is based on the $n$-crossed modules becomes natural and simple when $n=3$ and that the corresponding 3-form and 4-form gauge fields have to be a Dirac spinor and a Lorentz scalar, respectively. Hence by using a Poincar\'e 4-group we naturally incorporate fermionic and scalar matter into the corresponding 4-connection. The internal symmetries can be included into the 4-group structure by using a 3-crossed module based on the $SL(2,\C) \times K$ group, so that for $K=U(1)\times SU(2) \times SU(3)$ we can include the Standard Model into this categorification scheme.}

\begin{document}

\maketitle

\section{Introduction}

The central mathematical idea for the construction of the Standard Model (SM) was the concept of a Lie group and the corresponding connection on the principal bundle, i.e. the gauge symmetry. The SM group uniquely fixes the SM forces (the strong, the weak and the electro-magnetic force), while the matter content is not restricted by the SM group. Namely, the SM matter fields, i.e. the scalars and the fermions, can be a priori in any representation of the SM group. The SM representation is determined from the experiments and it is given by
\begin{equation} \label{smr}
\begin{array}{lcl} 
\r_{SM} & \!\!\! = \!\!\! & \ds (1,2,-\frac{1}{2}) \oplus \sum^3_{i=1} (3,2,\frac{1}{6})_i\oplus (\bar 3, 1, \frac{1}{3})_i \oplus \\
 & & \ds (\bar 3, 1, -\frac{2}{3})_i \oplus (1,2,-\frac{1}{2})_i \oplus (1,1, 1)_i \oplus (1,1,0)_i\,, \\
\end{array}
\end{equation}
where $(m,n,q)$ are the irreps of the SM group $SU(3)\times SU(2) \times U(1)$, see \cite{SM} for a mathematical review, and $i$ denotes a generation.

The spacetime symmetry properties of the SM fields are determined by the universal cover of the Lorentz group $SL(2,\C)$, while gravity, as the fourth fundamental force in Nature, can be understood via the gauge symmetry principle for the Poincar\'e group, i.e. as the Cartan connection 
\begin{equation}
(\o^{ab}, e^a ) = (\o^{ab}{}_\m \rmd x^\m, e^a{}_\m \rmd x^\m ) \,,
\end{equation}
where $\o^{ab}$ and $e^a$ are the spin-connection and the tetrad 1-forms,  $a,b = 1,2,3,4$  and $x^\m \in (x,y,z,t)$ are the spacetime coordinates.

One then wonders if there is a mathematical structure in a 4-dimensional spacetime which can incorporate the matter fields with the gauge fields and explain why the matter fields appear as the scalar and the spinor representations of the universal cover of the Lorentz group, as well as, why the representation $\r_{SM}$ appears. Note that the superstring theory is an example of such a structure, but it requires a 10-dimensional spacetime \cite{ST}.

As far as the internal symmetry group is concerned, the first attempt to explain $\r_{SM}$  was via the grand unification (GUT), for a review see \cite{GUT}. Although the fermionic part of one generation of $\r_{SM}$ neatly fits into the spinorial irrep of $SO(10)$, the GUT approach is problematic because of the appearance of many new gauge fields, i.e. new forces, so that the problem of symmetry breaking is the central problem to solve. One can even incorporate the Lorentz group (more precisely, its universal cover) into the GUT group, so that the Cartan connection becomes a part of the GUT group connection, see \cite{LoGUT}. However, this leads to an even larger GUT group so that the problem of symmetry breaking becomes even more difficult.

As far as the spacetime symmetry group is concerned, although the Poincar\'e gauge symmetry accommodates gravity and coupling of fermions, it does not restrict the $SL(2,\C)$ representations for the matter fields, since the Cartan connection contains only the spin-connection and the tetrads. In this letter we will propose a generalization of the local Poincar\'e symmetry, which will be given by the concept of a 4-group, defined as a 3-crossed module \cite{AKU}, so that one can include spinor and scalar fields as the components of the generalized connection associated to the Poincar\'e 4-group. Since $n$-groups are special categories which generalize the notion of symmetry, see \cite{cat}, we also show that the SM group can be easily included into the 4-group structure.

\section{General Relativity and categorical groups}

In \cite{MV} it was pointed out that General Relativity (GR) can be reformulated as a constrained $2BF$ theory for the Poincar\'e 2-group, which is defined as the crossed module
\begin{equation}
{\realni}^4 \xrightarrow{\pa}  SL(2,\C) \,,\label{2pg}
\end{equation}
where the map $\pa$ is trivial ($\pa(\vec{v}) =1_{SL(2,\C)}$ for all $\vec{v}\in \R^4$) while $SL(2,\C)$ acts on $\R^4$ as the vector representation. The Poincar\'e group then appears as the group of 2-morphisms of the 2-category which is equivalent to the crossed module (\ref{2pg}).

The Einstein-Cartan (EC) action can be obtained by constraining the $2BF$ action
\begin{equation}
S_2 = \int_\cM \left( B^{ab} \wedge R_{ab} + e^a \wedge G_a \right)\,,\label{2pga}
\end{equation}
where $B^{ab}$ is a 2-form, $e^a$ is a tetrad,
\begin{equation}
R_{ab} = \rmd \o_{ab} + \o_a{}^c \wedge \o_{cb} \,,\quad G_a = \rmd \b_a + \o_a{}^b \wedge \b_b \,,\label{rgc}
\end{equation}
are the components of the 2-curvature associated to the 2-connection 
\begin{equation}
   {\cal A}_2 = (\o^{ab}, \b^a  ) = (\o^{ab}{}_\m \,\rmd x^\m, \b^a{}_{\m\n}\, \rmd x^\m \wedge \rmd x^\n )\,,
\end{equation}
and $\b$ is the connection 2-form. The 2-connection ${\cal A}_2$ represents a categorical generalization (categorification) of the spin-connection. 

The constraint that transforms $S_2$ into the EC action is given by
\begin{equation}
B^{ab} = \e^{abcd} e_c \wedge e_d \,, \label{sc}
\end{equation}
where $\e^{abcd}$ is the totally antisymmetric tensor for the Poincar\'e group \cite{MV}. The constraint (\ref{sc}) is also known as the simplicity constraint.

The Poincar\'e 2-group structure does not restrict the matter representations, and in \cite{RV} it was pointed out that a Poincar\'e 3-group defined as a 2-crossed module based on the following group complex
\begin{equation}
   {\kompleksni}^4 \uto{\pa'} \,\,{\realni}^4   \uto{\pa} SL(2,\C) \,, \label{3pg}
\end{equation}
where the maps $\pa$ and $\pa'$ are trivial while $SL(2,\C)$ acts as the vector and the Dirac representation on $\R^4$ and $\C^4$, naturally gives the Dirac equation for the corresponding spinor field. 

Namely, one can associate the 3-connection
\begin{equation} \label{3con}
\begin{array}{c}
{\cal A}_3 = (\o^{ab}, \b^a , \G^\a ) = \\
(\o^{ab}{}_\m \, \rmd x^\m, \b^a{}_{\m\n} \, \rmd x^\m \wedge \rmd x^\n , \G^\a{}_{\m\n\r} \, \rmd x^\m \wedge \rmd x^\n \wedge \rmd x^\r )\,, \\
\end{array}
\end{equation}
to the 2-crossed module (\ref{3pg}), where $\G$ is a spinorial 3-form. The corresponding $3BF$ action is then given by
\begin{equation}
S_3 = \int_\cM \left( B^{ab} \wedge R_{ab} + e^a \wedge G_a + D_\a \wedge H^\a \right) \,,
\end{equation}
where $D_\a$ are 0-forms, while 
\begin{equation}
\quad H^\a = \rmd \G^\a + \o^\a{}_\b \wedge \G^\b \,,\quad \o^\a{}_\b = (\g_{a} )^\a{}_\d (\g_{b})^\d{}_\b \, \o^{ab} \,,\label{hc}
\end{equation}
is the curvature 4-form for $\G$, where $\g_a$ are the gamma matrices. 

The action $S_3$ can be converted into the EC action coupled to a Dirac fermion $\psi^\a$ by using the constraints 
\begin{equation}
\G^\a =\e_{abcd} \,e^a \wedge e^b \wedge e^c \, (\g^d)^\a{}_\b \psi^\b \,,\quad D_\a = \psi_\a \,, \label{3pdc}
\end{equation}
together with the simplicity constraint (\ref{sc}).

In order to obtain the complete EC Dirac action, one also has to add the spin-torsion and the mass term to $S_3$
\begin{equation} \label{stmdejstvo}
S_{Tm} =\int_\cM  \e_{abcd}\, e^a \wedge e^b \wedge \b^c \, \bar\psi \g^d \psi +\int_\cM |e| \, m\bar\psi \psi \, \rmd^4 x  \,,
\end{equation}
where $|e| =|\det (e^a{}_\m)|$.

However, if we want to associate some other $SL(2,\C)$ representation to the 3-form $\G$, then the task of obtaining the corresponding matter-field equation of motion becomes non-trivial,  since it is not easy to find the constraints for the corresponding $3BF$ action. For example, in the case of a real scalar field $\f$, one has a 2-crossed module
\begin{equation}
{\realni} \xrightarrow{\pa'}  {\realni}^4 \,\,\uto{\pa}  SL(2,\C) \,,\label{3pg2}
\end{equation}
and the constraints are given by
\begin{equation} \label{3psc}
\begin{array}{c}
  D=\f\,, \qquad \G = h_{abc} \,e^a \wedge e^b \wedge e^c \,, \\
  h_{abc} \e^{cdef} e_d \wedge e_e \wedge e_f =  \, e_a \wedge e_b  \wedge \rmd \f  \,, \\
\end{array}
\end{equation}
plus the simplicity constraint \cite{RV}. Since the Standard Model features the Higgs boson, it is important to be able to accommodate scalar fields in the formalism.

The last two constraints in (\ref{3psc}) are not easy to guess, so that one wonders: is it possible to resolve this difficulty by some higher categorical group? This can be done if we use a 4-group, which can be defined as a 3-crossed module \cite{AKU}.

Let us consider the following 3-crossed module (3CM)
\begin{equation}
   {\realni} \uto{\pa''} \,\,  {\kompleksni}^4 \uto{\pa'} \,\, {\realni}^4 \uto{\pa}  SL(2,\C) \,, \label{4pg}
\end{equation}
where $SL(2,\C)$ acts on $\R^4$, $\C^4$ and $\R$ as the vector, Dirac spinor and the scalar representation, while all other 3CM maps and actions are trivial. 

Note that a complex of Lie groups 
\begin{equation} U \uto{\pa''} W \uto{\pa'} V \uto{\pa} G \,,\label{g4cm}\end{equation} 
where $U,W$ and $V$ are Abelian groups corresponding to vector spaces of representations of $G$, is a 3-crossed module  if
\begin{enumerate}
\item $\pa'' \vec u = \vec 0_W$, $\pa' \vec w = \vec 0_V$, $\pa \vec v = 1_G$,
\item $g \triangleright \vec v = R_g \vec v$, $g \triangleright \vec w = R'_g \vec w$, $g \triangleright \vec u = R''_g \vec u$ (action of $G$ on $V, W$ and $U$),
\item $\vec v \triangleright' \vec w = \vec w$, $\vec v \triangleright' \vec u = \vec u$, $\vec w \triangleright'' \vec u = \vec u$ (action of $V$ on $W$ and $U$ and action of $W$ on $U$),
\item $\{\vec v , \vec v' \}_1 = \vec 0_W$, $\{\vec w , \vec w' \}_2 = \vec 0_U$, $\{\vec v , \vec w' \}_3 = \vec 0_U$, $\{\vec w , \vec v' \}_4 = \vec 0_U$ (The Peiffer maps $V\times V \to W$, $W\times W \to U$, $V\times W \to U$ and $W\times V \to U$ ).
\end{enumerate}

Given the Poincar\'e 4-group  (\ref{4pg}), we can construct the corresponding  4-connection as a collection of $p$-forms, $p=1,2,3,4$,
\begin{equation} {\cal A}_4 = (\o^{ab}, \b^a , \G^\a , \d )\,. \label{4con}\end{equation}
One can also promote ${\cal A}_4$ into a Lie-algebra-valued 4-connection by defining
\begin{equation} \hat{\cal A}_4 = (\o^{ab} J_{ab}, \b^a P_a , \G^\a Y_\a , \d X )\,, \label{h4con}\end{equation}
where $J,P,Y$ and $X$ are the generators of the Lie algebras for $SL(2,\C), \R^4, \C^4$ and $\R$ Lie groups.

Note that the 4-form $\d$ can be written as
\begin{equation}
\d = f (x,y,z,t)\, \rmd x \wedge \rmd y \wedge \rmd z \wedge \rmd t \,. \label{delta4form}
\end{equation}
Since $f$ is a scalar density, we will write $f = |e|\f$ and define the corresponding 1-form curvature as 
\begin{equation}
J = \rmd \f \,.
\end{equation}
Note that $\phi$ transforms as a $0$-form, i.e., as a scalar field, and is dual to the $4$-form $\delta$.
Then the 4-curvature for the 4-connection (\ref{4con}) will be given by
\begin{equation} {\cal F}_4 = (R^{ab}, G^a , H^\a , J ) \,,\end{equation}
where the $R$, $G$ and $H$ curvatures are given by (\ref{rgc}) and (\ref{hc}).

The $4BF$ action is then given by
\begin{equation}
S_4 = \int_\cM \left( B^{ab} \wedge R_{ab} + e^a \wedge G_a + \psi^\a \wedge H_\a + E \wedge J \right)\,,
\end{equation}
where $E$ is a 3-form. The EC action coupled to a Dirac and a scalar field is then obtained by imposing the constraints (\ref{sc}), (\ref{3pdc}) and
\begin{equation}
E_{\m\n\r} = |e|\,\e_{\m\n\r\s}  g^{\s\l}\pa_\l \f \,,\label{4psc}
\end{equation}
where $g^{\s\l}$ is the inverse metric of $g_{\m\n} = \eta_{ab} e^a{}_\m e^b{}_\n$ and $\e_{\m\n\r\s}$ is the Levi-Civita symbol. Note that now the scalar-field constraints are more natural and simpler than in the 2-crossed module case.

The complete EC action is then obtained by adding the fermion mass and the spin-torsion terms (\ref{stmdejstvo}) to $S_4$, as well as the scalar-field potential energy,
\begin{equation}
S_{V} = \int_\cM |e|  V(\f) \, \rmd^4 x \,. \label{sfpoten}
\end{equation}
Here $V(\f)$ is the potential for the scalar field. For the purpose of spontaneous symmetry breaking and the Higgs mechanism, one can introduce a doublet of complex scalar fields, and choose the standard Mexican hat potential,
\begin{equation}
V(\phi,\phi^\dag) = \lambda (\phi^\dag \phi - v^2)^2\,, \label{MexicanHat}
\end{equation}
where $\lambda$ is the quartic self-coupling of the scalar field, $v$ is the vacuum expectation value, and
\begin{equation}
\phi = \left( \begin{array}{c} \phi_+ \\ \phi_0 \\ \end{array} \right) \in \kompleksni^2 \label{HiggsDoublet}
\end{equation}
is the doublet of complex scalar fields. In order to accommodate a doublet of complex scalar fields, the first group $\realni$ in the 3CM chain complex (\ref{4pg}) should be substituted by $\kompleksni^2$, giving
\begin{equation}
   {\kompleksni^2} \uto{\pa''} \,\,  {\kompleksni}^4 \uto{\pa'} \,\, {\realni}^4 \uto{\pa}  SL(2,\C) \,. \label{4pgh}
\end{equation}
This choice of the 3CM will give rise to the complex doublet of the connection $4$-forms $\delta$ (see (\ref{delta4form})), whose dual will then be a doublet of $0$-forms (\ref{HiggsDoublet}).

\section{Standard Model and categorical groups}

The Poincar\'e 4-group (\ref{4pg}) can be easily modified in order to include the internal symmetries. Let us consider a 3-crossed module (\ref{g4cm}) given by
\begin{equation}
 \C^r \uto{\pa''}  \C^{2s'+2s''} \uto{\pa'} \R^4 \uto{\pa} \,\, SL(2,\C) \times K \,, \label{sm4g}
\end{equation}
where $K$ is a compact Lie group and $\C^{s'}$ is a vector space for a representation of $K$ for the left-handed fermions and  $\C^{s''}$ is a vector space for a representation of $K$ for the right-handed fermions. The left/right-handed fermions are described by the 2-component Weyl spinors corresponding to the $SL(2,\C)$ irreps $(\frac{1}{2}, 0)$ and $(0,\frac{1}{2})$  so that
\begin{equation}  \C^2 \otimes \C^{s'} \oplus \C^2 \otimes \C^{s''} \cong \C^{2s' + 2s''} \,.\end{equation}

The 4-connection which corresponds to (\ref{sm4g}) is given by a collection of $p$-forms, where $p=1,2,3,4$, and they can take values in the corresponding Lie algebras, so that
 \begin{equation} \hat{\cal A}_4 = (\o^{ab} J_{ab} + A^k T_k, \b^a P_a, \G^\a_j Y^j_\a , \d_i X^i)\,. \label{4aSM}\end{equation}
Here $T$, $Y$ and $X$ denote the generators of the Lie algebras for $K$, $\C^{2(s' + s'')}$ and $\C^r$ Lie groups. The 4-curvature for (\ref{4aSM}) will be given by
 \begin{equation} \hat{\cal F}_4 = (R^{ab} J_{ab} + F^k T_k, G^a P_a, H^\a_j Y^j_\a , J_i X^i)\,, \label{4fSM}\end{equation}
where 
\begin{equation}
\begin{array}{c}
\ds F^k T_k = \rmd A + \frac{1}{2}[A \wedge A] \,,\quad A=A^k T_k \,,\\
\ds H_j^\a = \rmd \G_j^\a + \o^\a{}_\b \wedge \G_j^\b \,, \quad J_i = d\f_i  \,. \\
\end{array}
\end{equation}

The $4BF$ action is then given by
\begin{equation} \label{yms4}
\begin{array}{lcl}
S_{4YM} & = & \ds \int_\cM \left( B^{ab} \wedge R_{ab} +B^{k} \wedge F_{k} + e^a \wedge G_a \right. \\
 & & \ds \hphantom{mmm} + \left. \psi^\a_j \wedge H_\a^j + E^i \wedge J_i \right)\,, \\
\end{array}
\end{equation}
where $B^k$ are 2-forms, $\psi_j^\a$ are 0-forms and $E^i$ are 3-forms.

The SM action coupled to GR is then obtained by using $K = SU(3) \times SU(2) \times U(1)$ and by constraining the $4BF$ action (\ref{yms4}) with the constraints (\ref{sc}) and with copies of the constraints (\ref{4psc}) and (\ref{3pdc}) for each $i$ and $j$. 

One also has to add to $S_{4YM}$ the potential terms quadratic in $B^k$ 
\begin{equation}
S_{YMP} = \int_\cM g^{\m\n} g^{\r\s} B^k{}_{\m\r} B_{k\n\s} \, \rmd^4 x \,,
\end{equation}
in order to obtain the Yang-Mills action, as well as the potential, the torsion and the Yukawa coupling terms for the matter fields $\f_i$ and $\psi_j^\a$.

The number of SM scalars $\f_i$ is determined by the Higgs doublet, see (\ref{smr}), hence $r=2$, similarly as in (\ref{4pgh}). Using this choice, and including into the action the scalar potential action (\ref{sfpoten}) with the choice (\ref{MexicanHat}), the Higgs mechanism applies in the standard way --- the $SU(2)\times U(1)$ subgroup of $K$ is spontaneously broken down to $U(1)_{em}$, three real-valued components in (\ref{HiggsDoublet}) are absorbed by the three gauge fields rendering them massive, while the fourth real component in (\ref{HiggsDoublet}) is interpreted as the Higgs field.

Finally, from (\ref{smr}) it follows that the number of SM fermions $\psi_j^\a$ is given by 
\begin{equation} 2s' + 2s'' = 2\cdot 12 \cdot 3 + 2\cdot 4 \cdot 3 = 96 \,,\end{equation} 
so that $s'=36$ and $s''=12$. The total number of fermionic components corresponds to 6 quarks plus 2 leptons, considered as Dirac spinors, for three generations, so that  $ 8\cdot 4 \cdot 3 = 96$.

\section{Conclusions}

We showed that a natural and simple categorification of GR based on $n$-crossed modules requires that $n=3$ and that the corresponding  2-form, 3-form and 4-form gauge fields have to be a vector, a Dirac spinor and a scalar, respectively.  Hence by using a categorical generalization of the Poincar\'e group, we naturally incorporate fermionic and  scalar matter  into the corresponding connection. The corresponding  Poincar\'e 4-group gauge field theory structure can be preserved by introducing the internal symmetries via the 3-crossed module (\ref{sm4g}), which can be considered as a categorical generalization of the $SL(2,\C) \times K$ symmetry group of SM.

Note that in the 3-group approach to SM \cite{RV}, one uses the 2-crossed module of the type
\begin{equation} U\times W \uto{\pa'} V \uto{\pa} G \,,  \end{equation}
which can be considered as a decategorification of the 3-crossed module (\ref{g4cm}). This is analogous to what happens in the case of pure gravity, where the Poincar\'e 2-group can be substitued by the Poincar\'e group, i.e. the $2BF$ action (\ref{2pga}) can be viewed as the $BF$ action for the Poincar\'e group, see \cite{MO}.

The 4-group (\ref{sm4g}) does not restrict the dimensions $r$, $s'$ and $s''$ so it would be interesting to explore if there exists another 4-group which is based on the group complex (\ref{sm4g}) but with different maps and actions such that $r$, $s'$ and $s''$ are related.

The ultimate goal would be to find a mathematical structure based on the 4-dimensional spacetime which can explain the dimensions $r$, $s'$ and $s''$. Our results suggest that categorical generalizations of groups can be useful for this goal, although some additional algebraic tools may be neccessary. See for example \cite{C}, where the McKay correspondence was proposed, or see \cite{B}, where the exceptional Jordan algebras were used. Whether the determination of $r$, $s'$ and $s''$  can be done classically or at the quantum level remains to be seen.

\acknowledgments
AM was supported by the FCT grant PTDC/MAT-PUR/31089/2017. MV was supported by the Ministry of Education, Science and Technological Development (MPNTR) of the Republic of Serbia, and by the Science Fund of the Republic of Serbia, Program DIASPORA, No. 6427195, SQ2020. The contents of this publication are the sole responsibility of the authors and can in no way be taken to reflect the views of the Science Fund of the Republic of Serbia.


\begin{thebibliography}{0}

\bibitem{SM}
  \Name{Baez J.~C. \and Huerta J.}
  \REVIEW{Bull. Amer. Math. Soc.}{47}{2010}{483}.

\bibitem{ST}
  \Name{Green M.~B., Schwarz J.~H. \and Witten E.}
  \Book{Superstring Theory: Volume 1, Introduction}
  \Publ{Cambridge University Press, Cambridge}
  \Year{1988}.
  
\bibitem{GUT}
  \Name{Georgi H.}
  \Book{Lie Algebras in Particle Physics: From Isospin to Unified Theories}
  \Publ{CRC Press, Boca Raton, Florida}
  \Year{1999}.

  
\bibitem{LoGUT}
  \Name{Lisi A.~G., Smolin L., \and Speziale S.}
  \REVIEW{Jour. Phys. A: Math. Theor.}{43}{2010}{445401}.

\bibitem{AKU}
  \Name{Arvasi Z., Kuzpinari T.~S. \and Uslu E.~{\"O}.}
  \REVIEW{Hom. Hom. Appl.}{11}{2009}{161}.

\bibitem{cat}
  \Name{Baez J.~C. \and Huerta J.}
  \REVIEW{Gen. Relativ. Gravit.}{43}{2011}{2335}.

\bibitem{MV}
  \Name{Mikovi{\'c} A. \and Vojinovi{\'c} M.}
  \REVIEW{Class. Quant. Grav.}{29}{2012}{165003}.

\bibitem{RV}
  \Name{Radenkovi{\'c} T. \and Vojinovi{\'c} M.}
  \REVIEW{JHEP}{10}{2019}{222}.

\bibitem{MO}
  \Name{Mikovi{\'c} A. \and Oliveira M.~A.}
  \REVIEW{Gen. Relativ. Gravit.}{47}{2015}{58}.

\bibitem{C}
  \Name{Crane L.}
  \REVIEW{Rev. Math. Phys.}{25}{2013}{1343005}.

\bibitem{B}
  \Name{Boyle L.}
  \REVIEW{\tt arXiv:2006.16265}{}{2020}{}.
 

\end{thebibliography}
\end{document}